\documentstyle[12pt]{article}  
\begin{document}    
\def\o{\over}
\def\Ar{\rightarrow}
\def\bar{\overline}
\def\r{\gamma}
\def\d{\delta}
\def\a{\alpha}
\def\b{\beta}
\def\n{\nu}
\def\m{\mu}
\def\k{\kappa}
\def\g{\gamma}
\def\r{\rho}
\def\s{\sigma}
\def\t{\tau}
\def\e{\epsilon}
\def\p{\pi}
\def\th{\theta}
\def\om{\omega}
\def\vp{{\varphi}}
\def\Re{{\rm Re}}
\def\Im{{\rm Im}}
\def\ra{\rightarrow}
\def\ti{\tilde}
\def\bar{\overline}
\def\l{\lambda}
\def\G{{\rm GeV}}
\def\M{{\rm MeV}}
\def\eV{{\rm eV}}
\setcounter{page}{1}
\thispagestyle{empty}
\topskip 1 cm
\begin{flushright}    
{EHU-99-1, \ January  1999}\\
\end{flushright}   
\vskip 1 cm
\centerline{\Large \bf Quark-Lepton Mass Matrices}
\vskip 0.5 cm
\centerline{\Large\bf  with $U(1)\times Z_2 \times Z_2'$ Flavor Symmetry}
\vskip 2 cm
\centerline{{\large \bf Morimitsu TANIMOTO}
  \footnote{E-mail address: tanimoto@edserv.ed.ehime-u.ac.jp} }
\vskip 1 cm
 \centerline{ \it{Science Education Laboratory, Ehime University, 
 790-8577 Matsuyama, JAPAN}}
\vskip 3 cm
\centerline{\Large \bf Abstract}
\vskip 1 cm
  The $U(1)$ flavor symmetry explains  the large mixing  of  neutrinos while 
  it  leads to the  unique texture for the quark mass matrices.
  It is remarked  that  $U(1)$ 
 symmetric mass matrices have the phenomenological defects.
In the quark sector, the mixing $V_{ub}$ is predicted to be  large
compared with the expected value $\l^4$ at the GUT scale.
 In the lepton sector,    $U(1)$ charges, which    give 
 a large mixing  in the neutrino sector,   also lead to
    the large one  in the charged lepton sector.
   In the viewpoint of the flavor symmetry,
 this is an unpleasant feature because the neutrino mass hierarchy is 
  determined only by  unknown   coefficients of ${\cal O}(1)$, and
    the near-maximal flavor  mixing
   is not guaranteed in the case of  both large  angle rotations.
  These defects  disappear  by introducing
 additional discrete symmetries   $Z_2\times Z_2'$.
 The $U(1)\times Z_2\times Z_2'$
 quark-lepton mass matrices are presented by  taking account of the
 recent data of  atmospheric neutrinos and  solar neutrinos. 

\vskip 1 cm

\newpage
\baselineskip=24 pt
\topskip 0 cm
\section{Introduction}

The standard model (SM)  has still  unexplained features such as      
      the quark-lepton mass spectra and the flavor mixings.
  Mixings of the quark sector (CKM matrix) \cite{KM} seem also to have an 
hierarchical structure.      
Those features may provide an important basis for a new physics beyond
the SM.                  
 On the other hand,  the flavor mixing of the lepton sector, so called
 MNS mixing matrix \cite{MNS} 
 is still ambiguous although  neutrino oscillation experiments  can  provide 
  information of the fundamental  property of neutrinos.  
  In these years, there is growing  experimental evidences of  neutrino
  oscillations.
  The exciting one is the atmospheric neutrino deficit \cite{Atm1}\cite{Atm2}
    as well as the solar neutrino deficit \cite{solar}\cite{SKamsolar}.
	   Super-Kamiokande \cite{SKam}  presented the near-maximal
     neutrino flavor oscillation in  atmospheric neutrinos.
 Furthermore a new stage is represented by the long baseline(LBL) neutrino oscillation
  experiments \cite{CHOOZ}\cite{K2K}\cite{MINOS}\cite{ICARUS}
 to confirm the large neutrino flavor oscillation.
  Since the CHOOZ result \cite{CHOOZ}
 excludes the large neutrino oscillation of $\nu_\mu \Ar \nu_e$ as far as
 $\Delta m^2 \geq 9\times 10^{-4} \eV^2$, 
 the large mixing between  $\nu_\m$ and $\nu_\tau$  is a reasonable interpretation
  for the atmospheric $\nu_\mu$ deficit.
  
  It will be important to understand why  there is the large flavor mixing in the
lepton sector
  in contrast to small mixings in the quark sector.
  Is there a possible flavor symmetry providing a large mixing angle
  in the  lepton mass matrices, which are consistent with the  quark ones?
  There is a simple explanation of the large mixing by the $U(1)$ flavor symmetry 
  \cite{U1}. 
  The $U(1)$ flavor symmetry leads to the  unique texture for the quark mass matrices
 \cite{Ramond}.
  However, it is remarked  that  $U(1)$ 
symmetric quark-lepton mass matrices have phenomenological defects.
 These defects  disappear  by introducing
 additional discrete symmetries   $Z_2\times Z_2'$.
In this paper, we discuss   phenomenological defects of 
  $U(1)$ symmetric mass matrices and present   $U(1)\times Z_2\times Z_2'$
quark-lepton mass matrices.

 Our approach  is to assume that oscillations need only account for the
 atmospheric and solar neutrino data.
 Since the result of LSND \cite{LSND} awaits confirmation by the KARMEN experiment
 \cite{KARMEN},
  we do not take into consideration the LSND data.
   Our starting point of   neutrino masses and  mixings is
   the atmospheric  $\nu_\m \Ar \nu_\tau$ oscillation with 
 \begin{equation}
  \Delta m^2_{\rm atm}=  5\times 10^{-4} - 6\times 10^{-3} \eV^2 \ , \qquad\quad
 \sin^2 2\th_{\rm atm} \geq 0.8 \ .
  \label{atm}
 \end{equation}
\noindent 
For the solar neutrino $\nu_e \Ar \nu_e$ oscillation,  
 some  solutions are still available \cite{BKS}.
In our paper, we take the small angle solution of MSW \cite{MSW}:
   \begin{equation}         
 \Delta m^2_\odot\simeq 5.4 \times 10^{-6} \eV^2 \ , 
\qquad\quad  \sin^2 2\theta_{\odot}        \simeq 6.0 \times 10^{-3}    \ .
   \end{equation}      
\noindent 
Since $\Delta m^2_{\rm atm}\gg \Delta m^2_\odot$, 
 we can take  neutrino masses as  
 \begin{equation}
  m_3\gg m_2\geq m_1 \ , 
 \label{mass1}
 \end{equation}
\noindent or
 \begin{equation}
    m_3\simeq m_2\simeq m_1 \ .
 \label{mass2}
 \end{equation}
\noindent
In the hierarchy  case of eq.(\ref{mass1}), the neutrino mass ratio of 
$m_2$ and $m_3$ is
 \begin{equation}
\frac{m_{\n_2}}{m_{\n_3}}={1\o 10} -  {1\o 30}\simeq \l^2  \ ,  
\label{mass3}
\end{equation}
\noindent
where $\l=0.22$. This ratio
 gives a strong constraint for the $Z_m$ symmetry \cite{Grossman}.
On the other hand, in the quasi-degenerate  case of eq.(\ref{mass2}),
the structure of the neutrino mass matrix is  restricted 
in the framework of the  $U(1)$ flavor  symmetry \cite{Pet}.

  \section{Phenomenology of $U(1)$ Flavor Symmetry}
  
  We start with discussing 
  well known regularities of the fermion mass ratios and the CKM matrix elements.
  The fermion mass ratios  at the GUT scale are given
 in terms of $\l\simeq 0.22$  \cite{Koide}  as follows:
  \begin{equation}
   {m_c\o m_t}\sim \l^4 \ ,\ \   {m_u\o m_t}\sim \l^8 \ , \ \
   {m_s\o m_b}\sim \l^2 \ , \ \   {m_d\o m_b}\sim \l^4 \ , \ \
  {m_\mu\o m_\tau}\sim \l^2 \ ,\ \    {m_e\o m_\tau}\sim \l^4 \ ,
  \end{equation}
\noindent for the internal mass hierarchy and
  \begin{equation}
    {m_b\o m_t}\sim \l^3 \ ,  \qquad  {m_b\o m_\tau}\sim 1 \ ,
  \end{equation}
\noindent  for the intrafamily hierarchy.
The CKM matrix elements at the GUT scale are 
\begin{equation}
   V_{us}=\l , \    V_{cb}=0.03-0.036\simeq \l^2 ,  \
   V_{ub}=0.0015- 0.004\simeq \l^4 ,  \ 
  \left |{V_{ub}\o V_{cb}}\right |= {1\o 4}\l - {1\o 2}\l , 
\label{CKM}
\end{equation}
\noindent which have been derived by using RGE's of the minimal SUSY model \cite{Koide}.
   In order to get these desirable relations of down quark masses and the CKM matrix 
elements, the natural down quark mass matrix   $M_{\rm d}$ is expressed
    in the hierarchy  base \cite{Ramond}:
 
 \begin{equation}         
   M_{\rm d} \simeq  \l^3 m_{0} \left (\matrix{    
                \l^\delta & \l^3 & \l^\eta \cr     
                    & \l^2 & \l^2 \cr    
                  &  & 1    \cr } \right )    \ ,
	\label{expd}
      \end{equation}        
 \noindent 
 where $\delta\geq 4$ and $\eta\geq 4$.
 On the other hand, the up quark mass matrix  $M_{\rm u}$ is 
 \begin{equation}         
   M_{\rm u} \simeq  m_{0} \left (\matrix{    
                \l^8 &  &  \cr     
                    & \l^4 & \l^\k \cr    
                  &  & 1    \cr } \right )    \ ,
 \label{expu}
      \end{equation}        
 \noindent 
 where $\k\geq 2$.
 The $U(1)$ flavor symmetry determines unknown entries.
 When we integrate out massive fermions the effective Yukawa couplings below the
  mass scale $\Lambda$ are of the form \cite{FrNi} \cite{Leurer}
  
  \begin{equation}         
    Q_i \bar d_j H_d \left (  {S\o \Lambda_d} \right )^{m_{ij}} +
    Q_i \bar u_j H_u \left (  {S\o \Lambda_u} \right )^{n_{ij}} + {\rm h.c.} \ ,
  \end{equation}
  \noindent
  where $S$ is a singlet scalar of the SM, which breaks the flavor symmetry 
spontaneously by a VEV $<S>$. For simplicity,
 we assume $\Lambda_d=\Lambda_u\equiv \Lambda$ and  $<S>/\Lambda=\l$ 
 to be like the Cabibbo angle.
  In non-supersymmetric models, powers of $S^\dagger$ should be allowed.
  However, since  this possibility is forbidden in the super-potential
 of the supersymmetric model, we ignore $S^\dagger$.
   
 Let us define $U(1)$ charges for doublet quarks $Q_i$ and singlets $\bar d_i$ and
 $\bar u_i$:
  \begin{equation}         
  ( Q_1, Q_2, Q_3 ) = (a_1, a_2, a_3), \ \ ( \bar d_1, \bar d_2, \bar d_3 ) =
  (x_1, x_2, x_3),
   \ \ ( \bar u_1, \bar u_2, \bar u_3 ) = (r_1, r_2, r_3) , 
  \end{equation}        
 \noindent 
 with $U(1)$ charges $h_d$, $h_u$ and $-1$ for the Higgs $H_d$, $H_u$
  and $S$, respectively.

 Then we get $m_{ij}=a_i+x_j+h_d$ and $n_{ij}=a_i+r_j+h_u \ (i,j=1,2,3)$.
 If  $m_{ij}$ and  $n_{ij}$ are negative, those entries are zeros.
 These rules together with experimental constraints of eqs.(\ref{expd}) and  (\ref{expu})
  uniquely fix  unknown entries of the quark mass matrix
   apart from  coefficients of  ${\cal O}(1)$ as follows:
  \begin{equation}         
   M_{\rm d} \simeq  \l^3 m_{0} \left (\matrix{    
                     \l^4 & \l^3 & \l^3 \cr     
                     \l^3   & \l^2 & \l^2 \cr    
                     \l & 1 & 1    \cr } \right )    \ ,
	\label{md}
      \end{equation}        
 \noindent 
 where $\delta=4$ and $\eta=3$ are fixed in eq.(\ref{expd}),  and 
  \begin{equation}         
   M_{\rm u} \simeq  m_{0} \left (\matrix{    
                  \l^8  &  \l^5 &  \l^3 \cr     
                  \l^7  &  \l^4 &  \l^2 \cr    
                  \l^5  &  \l^2 &   1   \cr } \right )    \ ,
   \label{mu}
      \end{equation}        
 where $\k=2$ is  fixed in eq.(\ref{expu}).
The second and third columns in $M_d$ are same due to $x_2=x_3$.
 We find a few undesirable features in these mass matrices in the viewpoint of 
 the quark mass matrix phenomenology.
 The first point is in (1-3) entries of $M_{\rm d}$ and $M_{\rm u}$.
 The successful relation between the CKM matrix elements and quark masses \cite{Hall}
   \begin{equation}         
       \left |\frac{V_{ub}}{V_{cb}} \right |\simeq \sqrt{\frac{m_u}{m_c}} \ ,
      \end{equation}    
 is derived by the assumption of zero texture in the (1-3) entry
 and the suppressed (3-2) entry.
  The ${\cal O}(\l^3)$ in the (1-3) entry, which is the same
order of the (1-2) entry,  spoils this relation considerably.
 This situation is easily understood 
  by taking into account  $V_{ub}=D_{13}+U^*_{21}D_{23}+U^*_{31}$, where
 $D$ and $U$ are  left-handed unitary matrices
  which diagonalize mass matrices $M_{\rm d}$ and $M_{\rm u}$, respectively.
   Mixings $D_{13}\sim \l^3$,  $U^*_{21}D_{23}\sim \l^3$ 
  and  $U^*_{31}\sim \l^3$
  are easily obtained from eqs.(\ref{md}) and (\ref{mu}) \cite{Hall}.
  Unless an accidental cancellation among $D_{13}$, $U^*_{21}D_{23}$
  and $U^*_{31}$ is realized,
  the experimentally consistent prediction  $V_{ub}\sim  \l^4$ is not expected.
  It may be usefull to comment on the effect of the  (3-2) entry.
 Even if the (1-3) entry is suppressed compared with $\l^3$, 
  $D_{13}\sim \l^3$ is still kept as far as  the 
 $M_d$ matrix has  such entries  as  $M_{d32}\simeq M_{d33}$
  due to  $M_{d12}\simeq \l^6$. The effect of the (3-2) entry
 is easily understood by the following formula:
 
 \begin{equation}         
    D_{13}\sim {M_{d13}\o M_{d33}} + {M_{d12} M_{d32}\o M_{d33}^2} \ .
 \end{equation}
 
  The second point is in the (1-2) entry in the up quark sector.
  The texture in eq.(\ref{mu}) gives a mixing  $U_{12}={\cal O}(\l)$
  for  the u-quark sector. 
  This does not guarantee a successful formula of \cite{mdms}
   \begin{equation}         
    V_{us}\simeq \sqrt{\frac{m_d}{m_s}} \ ,
  \end{equation}
\noindent
 because of the same order contribution from the up-quark sector.
 However,  $U_{12}={\cal O}(\l)$ may be harmless if 
 a coefficient smaller than 1 and a phase  are taken account.
  
  Let us consider the lepton sector.
  The effective Yukawa couplings below the mass scale $\Lambda$ are of the form
  \begin{equation}         
    L_i \bar \ell_j H_d \left (  {S\o \Lambda} \right )^{m_{ij}} +
    \frac{1}{M_R} L_i  L_j H_u H_u \left (  {S\o \Lambda} \right )^{n_{ij}} \ ,
 \label{effectiveL}
  \end{equation}
  \noindent
 where $M_R$ is a  relevant high mass scale, and 
 $U(1)$ charges for doublet leptons $L_i$ and singlets $\bar \ell_i$ are
 defined as
  \begin{equation}         
  ( L_1, L_2, L_3 ) = (\a_1, \a_2, \a_3), \qquad
  (\bar \ell_1, \bar \ell_2, \bar \ell_3)  = (\r_1, \r_2, \r_3)\ , 
  \end{equation}        
 \noindent  respectively.
  The charged lepton  mass matrix is  given apart from the constant mass factor as 
   \begin{equation}         
   M_{\rm \ell} \sim  \left (\matrix{    
          \l^{\a_1+\r_1}  &  \l^{\a_1+\r_2}  &  \l^{\a_1+\r_3}  \cr     
          \l^{\a_2+\r_1}   & \l^{\a_2+\r_2}  &  \l^{\a_2+\r_3}  \cr    
          \l^{\a_3+\r_1}  &  \l^{\a_3+\r_2}  &  \l^{\a_3+\r_3}  \cr } \right )  \ ,
  \label{me}   
   \end{equation}  
	and the effective left-handed Majorana neutrino mass matrix is   
	   \begin{equation}         
   M_{\rm \n} \sim  \left (\matrix{    
                  \l^{2\a_1}  &  \l^{\a_1+\a_2}   &  \l^{\a_1+\a_3}  \cr     
                  \l^{\a_1+\a_2}   &  \l^{2\a_2}  &  \l^{\a_2+\a_3} \cr    
                  \l^{\a_1+\a_3}   &  \l^{\a_2+\a_3}  & \l^{2\a_3}  \cr } \right ) \ ,
   \label{mn}     
 \end{equation}   
\noindent
where suppression factors due to the Higgs $U(1)$ charges $h_d$ and
 $h_d$ are omitted.
Hereafter, we omit the constant mass terms.
If the large mixing is required in the neutrino mixing matrix(MNS mixing matrix)
 \cite{MNS},
 $U(1)$ charges are determined.
The relation of $\a_2=\a_3$ makes the democratic submatrix in the (2-3) sector
of the neutrino mass matrix.  Then  the large  mixing is naturally obtained
 in the neutrino sector.  What happens in the charged lepton sector?
  The (2-i) and (3-i) entries $(i=1,2,3)$ are same ones due to $\a_2=\a_3$.
   In particular, the same value of (2-3) and (3-3) entries gives the large  mixing
 angle in the left-handed  charged lepton sector.
   Thus  $\a_2=\a_3$ essentially leads to large mixings in both neutrino sector and
    charged lepton sector.
   In the viewpoint of the flavor symmetry,
 this is an unpleasant feature because the neutrino mass hierarchy is 
  determined only by  unknown   coefficients of ${\cal O}(1)$.
 Moreover   the near-maximal MNS mixing
   is not guaranteed in the case of  both large angle rotations.
 Magnitudes of mixing angles and phases in both sector  must be tuned 
 if the experimental MNS mixing  in the (2-3) 
sector   is  the near-maximal mixing.
  
   Another way to get the large mixing with avoiding  this situation is to
  assume the relation $\a_2=-\a_3>0$ \cite{Pet}. Then  the neutrino mass hierarchy
turns to $m_3\simeq m_2\gg m_1$, which is contradict with the ones in eqs.(\ref{mass1})
 and (\ref{mass2}).
  Thus the $U(1)$ favor symmetry implies undesirable features in quark and lepton 
  sectors.
  The situation is not improved even if more $U(1)$ flavor symmetries
   such as $U(1)_1\times U(1)_2$
   or  $U(1)_1\times U(1)_2\times U(1)_3$  are introduced.

 To suppress $|V_{ub}|$ below $\l^3$, certain entries 
 in mass matrices have to be suppressed 
 relative to their naive values.
 This suppression is realized in the discrete symmetry $Z_m$, which is a subgroup
 of the $U(1)$ symmetry  as shown in  \cite{Leurer}.
 Anyway,
  in order to overcome these undesirable features of  quark-lepton mass matrices,
 one should consider beyond the $U(1)$  flavor symmetry.
 We discuss  the discrete symmetry $Z_m$ for both quark and lepton mass matrices.
 
\section{Quark-Lepton Mass Matrices with $Z_m$ Symmetry}
 
 A single $Z_m$ flavor symmetry is not helpful to improve above discussed
 undesirable features because certain entries in the mass matrix 
 cannot be suppressed relative to their naive values.
 Let us discuss the extended symmetry   $U(1)\times Z_m$.
 The effective Yukawa couplings of the lepton sector
 are given by extending eq.(\ref{effectiveL}) with new Higgs
  $S_1$ and  $S_2$ as follows \cite{Leurer}:
  \begin{equation}         
    L_i \bar \ell_j H_d \e_1^{m_{ij}} \e_2^{m'_{ij}} +
    \frac{1}{M_R} L_i  L_j H_u H_u \e_1^{n_{ij}} \e_2^{n'_{ij}}\ ,
  \end{equation}
  \noindent
 where $\e_1\equiv <S_1>/\Lambda$ and $\e_2\equiv <S_2>/\Lambda$ are assumed to be
  expressed in terms of $\l$.  
The $U(1)$ and $Z_m$ charges for doublet leptons $L_i$ and singlets 
 $\bar \ell_i$ are defined as
  \begin{eqnarray}         
  L_1(\a_1,\ \b_1) \ ,  \quad L_2(\a_2, \ \b_2) \ ,\quad  L_3(\a_3, \ \b_3) 
 \  ,\nonumber \\
  \bar \ell_1(\r_1,\ \s_1)\ ,\quad  \bar \ell_2(\r_2, \ \s_2 )\ ,
               \quad \bar \ell_3(\r_3, \ \s_3) \ , 
  \end{eqnarray}        
 \noindent 
 respectively.
In order to see  the neutrino mass hierarchy and the large mixing, we discuss the (2-3)
submatrices of the charged lepton and the left-handed Majorana neutrino mass matrices:   
      \begin{equation}         
   M_{\rm \ell} \sim \left (\matrix{    
     \e_1^{\a_2+\r_2}\e_2^{\b_2+\s_2} &\e_1^{\a_2+\r_3}\e_2^{\b_2+\s_3}  \cr     
     \e_1^{\a_3+\r_2}\e_2^{\b_3+\s_2} &\e_1^{\a_3+\r_3}\e_2^{\b_3+\s_3} } \right ) \ ,
      \end{equation}
\noindent and
  \begin{equation}         
   M_{\rm \n} \sim  \left (\matrix{    
                \e_1^{2\a_2}\e_2^{2\b_2}  &  \e_1^{\a_2+\a_3}\e_2^{\b_2+\b_3}   \cr     
       \e_1^{\a_2+\a_3}\e_2^{\b_2+\b_3}   &  \e_1^{2\a_3}\e_2^{2\b_3} } \right ) \ ,
      \end{equation}   
\noindent 
  respectively.
 If we take  $\a_2=\a_3$ for $U(1)$ charges  and $\b_1=\b_2$ for $Z_m$ charges,
 both  mixing angles of charged leptons and neutrinos
 could be large as discussed in eqs.(\ref{me}) and (\ref{mn}).
 However  only the mixing angle of the charged lepton could be large due to
 the  $Z_m$ symmetry  if we put 
 \begin{equation}
\b_2+\s_3=m \ , \qquad \quad
\e_1^{\a_2}= \e_1^{\a_3}\e_2^{\b_3+\s_3} \ , 
\label{condm}
\end{equation}
\noindent  with $\b_3+\s_3 \not= 0$.
 Then the neutrino mass matrix has a hierarchical structure.
By use of these two conditions, we get

 \begin{equation}         
   M_{\rm \n} \sim  \left (\matrix{    
     \e_2^{2m}  &  \e_2^{m}\cr \e_2^{m} &  1} \right )\e_1^{2\a_3}\e_2^{2\b_3}  \ .
 \end{equation}  
\noindent
 By taking $\e_2=\l$, we get the neutrino mass ratio
 $m_2: m_3= \l^{2m}:1$.  As seen in eq.(\ref{mass3}), the experimental
data suggest $m=1$, which is unfavorable for the $Z_m$ symmetry.
 However, as shown by Grossman, Nir and Shadmi \cite{Grossman},
 if the condition $2\b_2=m$ is added to eq.(\ref{condm}),
the neutrino mass matrix turns to
 \begin{equation}         
   M_{\rm \n} \sim  \left (\matrix{    
     \e_2^{m}  &  \e_2^{m}\cr \e_2^{m} &  1} \right )\e_1^{2\a_3}\e_2^{2\b_3}  \ ,
 \end{equation}  
\noindent
which leads to $m_2: m_3= \l^{m}:1$.
  It is remarked that the mass enhancement of $m_2$ is realized due to the $Z_m$ symmetry
\cite{Grossman}.
Only the $Z_2$ symmetry is consistent with the experimental mass ratio
of $m_2$ and $m_3$.

 After putting $U(1)$ charges  $\a_3=0$, $\a_2=1$, $\r_3=2$ and
  $Z_2$ charges $\b_3=0$, $\b_2=1$,  $\s_3=1$, which satisfy
 above conditions,  it is easily found that
 $\e_1=\l$ should be fixed as well as $\e_2=\l$  to get 
  $m_2: m_3= \l^{2}:1$ in the neutrino mass matrix.
 Therefore  we take $\e_1=\e_2=\l$ in following analyses.

 Let us study the quark sector in the $U(1)\times Z_2$ symmetry.
The effective Yukawa couplings of the quark  sector
are of the form
  \begin{equation}         
    Q_i \bar d_j H_d \e_1^{m_{ij}} \e_2^{m'_{ij}} +
    Q_i \bar u_j H_u \e_1^{n_{ij}} \e_2^{n'_{ij}}  \ .
  \end{equation}
\noindent
 The $U(1)$ and $Z_2$ charges for  doublet quarks $Q_i$, singlets
 $\bar d_i$  and   $\bar u_i$  are defined as
  \begin{eqnarray}         
  Q_1(a_1,\ b_1) \ ,  \quad Q_2(a_2, \ b_2) \ ,\quad  Q_3(a_3, \ b_3) 
 \  ,\nonumber \\
  \bar d_1(x_1,\ y_1)\ ,\quad  \bar d_2(x_2, \ y_2 )\ ,
               \quad \bar d_3(x_3, \ y_3) \ ,  \nonumber \\
  \bar u_1(r_1,\ s_1)\ ,\quad  \bar u_2(r_2, \ s_2 )\ ,
               \quad \bar u_3(r_3, \ s_3) \ , 
  \end{eqnarray}        
 \noindent 
 respectively.
 Then  we can express  quark mass matrices in terms of
 $U(1)$ and $Z_2$ charges apart from order one coefficients:
\begin{equation}         
   M_{\rm d} \sim  \left (\matrix{    
\e_1^{a_1+x_1}\e_2^{b_1+y_1} &\e_1^{a_1+x_2}\e_2^{b_1+y_2} & \e_1^{a_1+x_3}\e_2^{b_1+y_3}               \cr
\e_1^{a_2+x_1}\e_2^{b_2+y_1} &\e_1^{a_2+x_2}\e_2^{b_2+y_2} & \e_1^{a_2+x_3}\e_2^{b_2+y_3}              \cr
\e_1^{x_1}\e_2^{y_1} & \e_1^{x_2}\e_2^{y_2} & \e_1^{x_3}\e_2^{y_3}\cr
                   } \right )    \ ,
	\label{qd1}
      \end{equation}        
\noindent and 
 \begin{equation}         
   M_{\rm u} \sim  \left (\matrix{    
\e_1^{a_1+r_1}\e_2^{b_1+s_1} &\e_1^{a_1+r_2}\e_2^{b_1+s_2} &
 \e_1^{a_1+r_3}\e_2^{b_1+s_3}\cr
\e_1^{a_2+r_1}\e_2^{b_2+s_1} &\e_1^{a_2+r_2}\e_2^{b_2+s_2} & 
\e_1^{a_2+r_3}\e_2^{b_2+s_3}\cr
\e_1^{r_1}\e_2^{s_1} & \e_1^{r_2}\e_2^{s_2} & \e_1^{r_3}\e_2^{s_3}\cr
                   } \right )    \ ,
 \label{qu1}
      \end{equation}        
 \noindent  where $a_3$ and $b_3$ are set to be zero without loss of generality. 

In order to avoid a phenomenological defect $D_{13}\sim \l^3$,
we search $U(1)$ and $Z_2$ charges leading to  $M_{d32}\sim\l^5$  and
  $M_{d13}\leq \l^7$
under four conditions
\begin{equation}
  M_{d33}=\l^3,\quad M_{d23} =\l^5,\quad   M_{d22} =\l^5,\quad  M_{d12} =\l^6.
\label{mdcond}
\end{equation}
We get only two  solutions to satisfy these four  conditions and $M_{d32}\sim\l^5$
as
\begin{eqnarray} 
(x_3=3, \ \ y_3=0,\ \ x_2=4, \ \ y_2=1, \ a_2=1,\ b_2=1,  \ a_1=1,\ b_1=0) , 
\nonumber \\
(x_3=3, \ \ y_3=0,\ \ x_2=4, \ \ y_2=1, \ a_2=1,\ b_2=1,  \ a_1=2,\ b_1=1) .
\end{eqnarray}
\noindent
However we get 
$M_{d13}= \l^4$ and $M_{d13}= \l^6$ for each solution, respectively.
Thus  $M_{d13}\leq \l^7$ cannot be obtained  in the framework of the 
$U(1)\times Z_2$ flavor symmetry.
In conclusion,  the defect of $V_{ub}$ is still kept 
 although the defect in the lepton sector is removed.

If the $Z_m$ symmetry with $m\geq 3$ is  introduced,  the problem of $V_{ub}$
could be resolved as shown in ref.\cite{Leurer}. However
the neutrino mass ratio in eq.(\ref{mass3}) allows only $m=2$.  
In order to remove the defect of $V_{ub}$, we should proceed to
the minimal  extension 
 $U(1)\times Z_2\times Z_2'$ flavor symmetry.
 \footnote{We distinguish two $Z_2$ symmetries by using prime.}

\section{Mass matrices with  $U(1)\times Z_2\times Z_2'$ Symmetry}

 In the  $U(1)\times Z_2\times Z_2'$ symmetry,
 the effective Yukawa couplings of the lepton sector
 are given by extending eq.(\ref{effectiveL}) with three Higgs
  $S_1$, $S_2$ and $S_3$ as follows:
  \begin{equation}         
    L_i \bar \ell_j H_d \e_1^{m_{ij}} \e_2^{m'_{ij}} \e_3^{m''_{ij}} +
    \frac{1}{M_R} L_i  L_j H_u H_u \e_1^{n_{ij}} \e_2^{n'_{ij}}  \e_3^{n''_{ij}}\ ,
  \end{equation}
  \noindent
 where a new parameter $\e_3\equiv <S_3>/\Lambda$ is introduced.
The $U(1)$, $Z_2$ and $Z_2'$ charges for doublet leptons $L_i$ and singlets 
 $\bar \ell_i$ are:

  \begin{eqnarray}         
  L_1(\a_1,\ \b_1, \ \g_1) \ ,  \quad L_2(\a_2, \ \b_2, \ \g_2) \ ,\quad
  L_3(\a_3, \ \b_3,\  \g_3) 
 \  ,\nonumber \\
  \bar \ell_1(\r_1,\ \s_1, \ \t_1)\ ,\quad  \bar \ell_2(\r_2, \ \s_2, \ \t_2)\ ,
               \quad \bar \ell_3(\r_3, \ \s_3, \ \t_3) \ , 
  \end{eqnarray}
\noindent respectively.
The charged lepton and the neutrino mass matrices are given as follows:

\begin{equation}         
   M_{\rm \ell} \sim  \left (\matrix{    
\e_1^{\a_1+\r_1}\e_2^{\b_1+\s_1} \e_3^{\g_1+\t_1} & 
\e_1^{\a_1+\r_2}\e_2^{\b_1+\s_2} \e_3^{\g_1+\t_2} & 
\e_1^{\a_1+\r_3}\e_2^{\b_1+\s_3} \e_3^{\g_1+\t_3}     \cr
\e_1^{\a_2+\r_1}\e_2^{\b_2+\s_1} \e_3^{\g_2+\t_1} & 
\e_1^{\a_2+\r_2}\e_2^{\b_2+\s_2} \e_3^{\g_2+\t_2} & 
\e_1^{\a_2+\r_3}\e_2^{\b_2+\s_3} \e_3^{\g_2+\t_3}     \cr
\e_1^{\a_3+\r_1}\e_2^{\b_3+\s_1} \e_3^{\g_3+\t_1} & 
\e_1^{\a_3+\r_2}\e_2^{\b_3+\s_2} \e_3^{\g_3+\t_2} & 
\e_1^{\a_3+\r_3}\e_2^{\b_3+\s_3} \e_3^{\g_3+\t_3}     \cr
                   } \right )    \ ,
	\label{L3}
\end{equation}        
\noindent and
 
\begin{equation}         
   M_{\rm \n} \sim  \left (\matrix{    
\e_1^{2\a_1}    \e_2^{2\b_1}     \e_3^{2\g_1} & 
\e_1^{\a_1+\a_2}\e_2^{\b_1+\b_2} \e_3^{\g_1+\g_2} & 
\e_1^{\a_1+\a_3}\e_2^{\b_1+\b_3} \e_3^{\g_1+\g_3}     \cr
\e_1^{\a_2+\a_1}\e_2^{\b_2+\b_1} \e_3^{\g_2+\g_1} & 
\e_1^{2\a_2}    \e_2^{2\b_2}     \e_3^{2\g_2} & 
\e_1^{\a_2+\a_3}\e_2^{\b_2+\b_3} \e_3^{\g_2+\g_3}     \cr
\e_1^{\a_3+\a_1}\e_2^{\b_3+\b_1} \e_3^{\g_3+\g_1} & 
\e_1^{\a_3+\a_2}\e_2^{\b_3+\a_2} \e_3^{\g_3+\g_2} & 
\e_1^{2\a_3}    \e_2^{2\b_3}     \e_3^{2\g_3}     \cr
                   } \right )    \ ,
	\label{N3}
\end{equation}        
\noindent where 
we can take $\a_3=0$, $\b_3=0$ and  $\g_3=0$   without loss of  generality. 	  
In order to get the large mixing angle of the (2-3) sector in the charged lepton mass
 matrix keeping the small mixing angle in the neutrino mass matrix,
 we take similar  conditions as  in eq.(\ref{condm}),
 either
\begin{equation}
\b_2+\s_3=2 \ , \qquad 
\e_1^{\a_2}\e_3^{\g_2+\t_3} =\e_2^{\s_3}\e_3^{\t_3} \ , 
\label{condm2}
\end{equation} 
\noindent   or
\begin{equation}
\g_2+\t_3=2 \ , \qquad 
\e_1^{\a_2}\e_2^{\b_2+\s_3} =\e_2^{\s_3}\e_3^{\t_3} \ . 
\label{condm3}
\end{equation} 
\noindent
 Then the neutrino mass matrix has a hierarchy structure.
 By taking  either $\e_1=\l$,  $\e_{2}=\l$ and $\a_2=1$
 in addition to eq.(\ref{condm2}) or
$\e_1=\l$,  $\e_{3}=\l$ and $\a_2=1$
 in addition to eq.(\ref{condm3}),
 we get the neutrino mass ratio
 $m_2: m_3= \l^{2}:1$.
In the first(second) case,   $\e_{3}$($\e_{2}$) is not nesessary to be $\l$. 
  However we take  $\e_1=\e_2=\e_3=\l$  for simplicity.
 We can fix  $\a_2, \ \b_2, \ \g_2$ and $\s_3$($\t_3$)
 by the condition eq.(\ref{condm2})(eq.(\ref{condm3})) and the neutrino
mass hierarchy as follows:
\begin{eqnarray} 
(\a_2=1, \quad \b_2=1,\quad  \g_2=0, \quad \s_3=1) , 
\nonumber \\
(\a_2=1, \quad \b_2=0,\quad \g_2=1, \quad \t_3=1) .
\end{eqnarray}
\noindent
Since two solutions are equivalent after interchange
 of $\e_2$ and $\e_3$, we consider only the solution of the
 first case hereafter.
The condition $M_{\ell 33}=\l^3$ gives two set of  $(\r_3, \ \t_3)$
\begin{equation} 
  (\r_3=2, \ \ \t_3=0),  \quad {\rm or } \quad    (\r_3=1, \ \ \t_3=1) \ .
\label{sol3}
\end{equation}
\noindent
 Furthermore conditions  $M_{\ell 22}=\l^5$ and $M_{\ell 32}=\l^5$   fix
\begin{equation} 
 (1)\   (\r_2=4, \ \s_2=1,  \ \t_2=0),  \quad (2)\  (\r_2=3, \ \s_2=1,  \ \t_2=1)  \ .
\label{sol2}
\end{equation}
At the last step, we can fix  charges by using conditions
$M_{\ell 11}\sim\l^7$, $M_{\ell 12}\leq \l^7$ and  $M_{\ell 21}M_{\ell 12}\leq \l^{12}$.
For the solution (1) in eq.(\ref{sol2})
 we get two sets of charges:
\begin{equation} 
  (A)\ (\a_1=1, \ \b_1=0,  \ \g_1=1), \quad  (\r_1=4, \ \s_1=1,  \ \t_1=0)  \ ,
\end{equation}
\begin{equation} 
  (B)\ (\a_1=1, \ \b_1=0,  \ \g_1=1), \quad  (\r_1=5, \ \s_1=0,  \ \t_1=0)  \ ,
\end{equation}
\noindent
and for the solution (2):
\begin{equation} 
  (C)\ (\a_1=2, \ \b_1=0,  \ \g_1=0), \quad  (\r_1=4, \ \s_1=1,  \ \t_1=0)  \ ,
\end{equation}
\begin{equation} 
  (D)\ (\a_1=2, \ \b_1=0,  \ \g_1=0), \quad  (\r_1=5, \ \s_1=0,  \ \t_1=0)  \ .
\end{equation}
\noindent
For cases (A) and (B), 
 $(\r_3=1, \ \t_3=1)$ in eq.(\ref{sol3}) leads to 
   $M_{\ell 13}\sim\l^3$, which gives the democratic mixing
 for left-handed charged leptons.
For other cases,  $M_{\ell 13}\sim\l^5$ is predicted.

We show  charged lepton mass matrices and  neutrino ones
 for each case:

\begin{equation} 
(A)\    M_{\rm \ell} \sim  \left (\matrix{    
    \l^7 & \l^7 & \l^{5(3)} \cr
    \l^5 & \l^5 & \l^3 \cr
    \l^5 & \l^5 & \l^3 \cr
                   } \right )    \ ,
\qquad
 M_{\rm \n} \sim  \left (\matrix{    
    \l^2 & \l^4 & \l^2 \cr
    \l^4 & \l^2 & \l^2 \cr
    \l^2 & \l^2 & 1 \cr
                   } \right )    \ ,
\end{equation}
\begin{equation} 
(B)\    M_{\rm \ell} \sim  \left (\matrix{    
    \l^7 & \l^7 & \l^{5(3)} \cr
    \l^7 & \l^5 & \l^3 \cr
    \l^5 & \l^5 & \l^3 \cr
                   } \right )    \ ,
\qquad
 M_{\rm \n} \sim  \left (\matrix{    
    \l^2 & \l^4 & \l^2 \cr
    \l^4 & \l^2 & \l^2 \cr
    \l^2 & \l^2 & 1 \cr
                   } \right )    \ ,
\end{equation}
\begin{equation} 
(C)\    M_{\rm \ell} \sim  \left (\matrix{    
    \l^7 & \l^7 & \l^5 \cr
    \l^5 & \l^5 & \l^3 \cr
    \l^5 & \l^5 & \l^3 \cr
                   } \right )    \ ,
\qquad
 M_{\rm \n} \sim  \left (\matrix{    
    \l^4 & \l^4 & \l^2 \cr
    \l^4 & \l^2 & \l^2 \cr
    \l^2 & \l^2 & 1 \cr
                   } \right )    \ ,
\end{equation}
\begin{equation} 
(D)\    M_{\rm \ell} \sim  \left (\matrix{    
    \l^7 & \l^7 & \l^5 \cr
    \l^7 & \l^5 & \l^3 \cr
    \l^5 & \l^5 & \l^3 \cr
                   } \right )    \ ,
\qquad
 M_{\rm \n} \sim  \left (\matrix{    
    \l^4 & \l^4 & \l^2 \cr
    \l^4 & \l^2 & \l^2 \cr
    \l^2 & \l^2 & 1 \cr
                   } \right )    \ .
\end{equation}
 The neutrino mass hierarchy is
 $m_3:m_2:m_1=1:\l^2:\l^2$ for cases (A) and (B),
 and   $m_3:m_2:m_1=1:\l^2:\l^4$ for cases (C) and (D).
 In these four solutions,
 we expect  neutrino mixings $U_{e2}\sim \l^2$ and $U_{e3}\sim\l^2$
 as well as $U_{\m2}\simeq 1/\sqrt{2}$.
It may be useful to comment on that $U_{e2}$ could be also  large  for cases
(A) and (B) if coefficients of $M_{\n 11}$ and $M_{\n 22}$ close each other.
However we do not expect such a difference of ${\cal O}(\l^4)$ between $M_{\n 11}$ and 
 $M_{\n 22}$.  In these cases, $M_{\ell 13}\simeq \l^3$ also provides the possibility
to give a large $U_{e2}$.

 Let us discuss  quark mass matrices in the $U(1)\times Z_2 \times Z_2'$
 flavor symmetry.
The effective Yukawa couplings of the quark  sector
are of the form
  \begin{equation}         
    Q_i \bar d_j H_d \e_1^{m_{ij}} \e_2^{m'_{ij}} \e_3^{m''_{ij}}  +
    Q_i \bar u_j H_u \e_1^{n_{ij}} \e_2^{n'_{ij}} \e_3^{n''_{ij}} \ .
  \end{equation}
\noindent
 The $U(1)$, $Z_2$ and $Z_2'$  charges for doublet quarks $Q_i$, singlets
 $\bar d_i$  and   $\bar u_i$  are:
  \begin{eqnarray}         
  Q_1(a_1,\ b_1, \ c_1) \ ,  \quad Q_2(a_2, \ b_2, \ c_2) \ ,\quad  
  Q_3(a_3,\ b_3, \ c_3) 
 \  ,\nonumber \\
  \bar d_1(x_1,\ y_1,\ z_1)\ ,\quad  \bar d_2(x_2, \ y_2, \ z_2)\ ,
               \quad \bar d_3(x_3, \ y_3, \ z_3) \ ,  \nonumber \\
  \bar u_1(r_1,\ s_1, \ t_1)\ ,\quad  \bar u_2(r_2, \ s_2, \  t_2)\ ,
               \quad \bar u_3(r_3, \ s_3, \ t_3) \ , 
  \end{eqnarray}        
 \noindent 
 respectively.
 Then  quark mass matrices are written as
\begin{equation}         
   M_{\rm d} \sim  \left (\matrix{    
\e_1^{a_1+x_1}\e_2^{b_1+y_1}\e_3^{c_1+z_1} & \e_1^{a_1+x_2}\e_2^{b_1+y_2}\e_3^{c_1+z_2}
 & \e_1^{a_1+x_3}\e_2^{b_1+y_3}\e_3^{c_1+z_3}    \cr
\e_1^{a_2+x_1}\e_2^{b_2+y_1}\e_3^{c_2+z_1} & \e_1^{a_2+x_2}\e_2^{b_2+y_2}\e_3^{c_2+z_2}
 & \e_1^{a_2+x_3}\e_2^{b_2+y_3}\e_3^{c_2+z_3}    \cr
\e_1^{x_1}\e_2^{y_1}\e_3^{z_1} & \e_1^{x_2}\e_2^{y_2}\e_3^{z_2}
 & \e_1^{x_3}\e_2^{y_3}\e_3^{z_3} \cr
                   } \right )    \ ,
	\label{qd2}
      \end{equation} 
\begin{equation}         
   M_{\rm u} \sim  \left (\matrix{    
\e_1^{a_1+r_1}\e_2^{b_1+s_1}\e_3^{c_1+t_1} & \e_1^{a_1+r_2}\e_2^{b_1+s_2}\e_3^{c_1+t_2}
 & \e_1^{a_1+r_3}\e_2^{b_1+s_3}\e_3^{c_1+t_3}    \cr
\e_1^{a_2+r_1}\e_2^{b_2+s_1}\e_3^{c_2+t_1} & \e_1^{a_2+r_2}\e_2^{b_2+s_2}\e_3^{c_2+t_2}
 & \e_1^{a_2+r_3}\e_2^{b_2+s_3}\e_3^{c_2+t_3}    \cr
\e_1^{r_1}\e_2^{s_1}\e_3^{t_1} & \e_1^{r_2}\e_2^{s_2}\e_3^{t_2}
 & 1 \cr
                   } \right )    \ ,
	\label{qu2}
      \end{equation}  
\noindent
 where we take $a_3=b_3=c_3=0$ and $r_3=s_3=t_3=0$ without loss of generality.
As shown in the previous section, relevant charges are obtained
 under four  conditions in eq.(\ref{mdcond}).
  At first, 
we search  $U(1)$ and $Z_2$ charges leading to  $M_{d32}\sim\l^5$
 and $M_{d33}\sim\l^3$.
The unique solution is 
  \begin{equation}         
  (x_3=3,  \ y_3=0, \ z_3=0),  \qquad  (x_2=3,  \ y_2=1, \ z_1=1),  
  \end{equation}
\noindent because  $x_3\leq 2$ and $x_2\geq 4$ always leads to  $M_{d12}\leq M_{d13}$,
 which is not consistent with the observed CKM matrix.
Next, we get two sets 
  \begin{equation}         
  (a_2=1,  \ b_2=1, \ c_2=0),   \qquad   (a_2=1,  \ b_2=0, \ c_2=1), 
  \end{equation}
\noindent
 by using  conditions  $M_{d22}\sim\l^5$ and $M_{d23}\sim\l^5$.
Now our concerned elements  $M_{d12}$ and $M_{d13}$ are
 \begin{equation}         
  M_{d12}=\e_1^{a_1+3}\e_2^{b_1+1}\e_3^{c_1+1} ,   \qquad 
  M_{d13}=\e_1^{a_1+3}\e_2^{b_1}\e_3^{c_1}.
\label{d12d13}
  \end{equation}
\noindent  By taking
\begin{equation}         
  a_1=3,  \quad  b_1=1, \quad  c_1=1 ,
  \end{equation}
\noindent  we get desirable elements  $M_{d12}=\l^6$ and $M_{d13}=\l^8$.
This is due to the $Z_2\times Z_2'$ symmetry as seen in eq.(\ref{d12d13}).
At the last step, the down quark mass ratio of $m_d$ and $m_s$ gives
\begin{equation}         
  x_1=4,  \quad y_1=1, \quad c_1=1. 
  \end{equation}
\noindent
The down quark mass matrix is uniquely determined as follows:
\begin{equation}  
M_{\rm d} \sim  \left (\matrix{    
    \l^7 & \l^6 & \l^8 \cr
    \l^6 & \l^5 & \l^5 \cr
    \l^6 & \l^5 & \l^3 \cr
                   } \right )    \ .
\end{equation}
 Let us consider the up quark sector.
By the condition  $M_{u22}=\l^4$, we get
\begin{equation}         
 (1)\ r_2=2,  \ s_2=0, \ t_2=0, \quad  (2)\ r_2=2,  \ s_2=1, \ t_2=1, \quad
 (3)\ r_2=3,  \ s_2=1, \ t_2=0. 
\end{equation}
\noindent
 Each solution gives
\begin{equation}         
 (1)\  M_{u12}=\l^7, \  M_{u32}=\l^2, \ \
 (2)\  M_{u12}=\l^5, \  M_{u32}=\l^4, \ \
 (3)\  M_{u12}=\l^7, \  M_{u32}=\l^4,
\label{u12}
\end{equation}
\noindent respectively.
Since  solutions (1) and (3) in eq.(\ref{u12}) lead to rather small 
$V_{ub}$ ($\simeq${\cal O}($\l^5$)), we choose the solution (2).

The condition  $M_{u11}=\l^8$ determines  $(r_1,  \ s_1, \ t_1)$ parameters
as follows: 
\begin{eqnarray}         
 (1)\ r_1=5,  \ s_1=1, \ t_1=1, \quad  (2)\ r_1=4,  \ s_1=0, \ t_1=1, \nonumber \\
 (3)\  r_1=4,  \ s_1=1, \ t_1=0. \quad (4)\  r_1=3,  \ s_1=0, \ t_1=0 ,
\end{eqnarray}
\noindent which lead to 
\begin{eqnarray}         
 (1)\  M_{u21}=\l^7, \  M_{u31}=\l^7, \quad
 (2)\  M_{u21}=\l^7, \  M_{u31}=\l^5, \nonumber \\
 (3)\  M_{u21}=\l^5, \  M_{u31}=\l^5, \quad
 (4)\  M_{u21}=\l^5, \  M_{u31}=\l^3,
\label{u31}
\end{eqnarray}
\noindent  respectively.
The solutions (3) and (4) are excluded because those  give the large $m_u$ 
due to  $M_{u12}\simeq M_{u21}\simeq \l^5$.
In conclusion,
the up quark mass matrix is given  as follows:
\begin{equation}  
M_{\rm u} \sim  \left (\matrix{    
    \l^8 & \l^5 & \l^5 \cr
    \l^7 & \l^4 & \l^2 \cr
   \l^{7(5)}   & \l^2 &  1 \cr
                   } \right )  \  .
\end{equation}


In our quark mass matrices,  (1-3) entries
are suppressed in both down and up quark sectors
due to the additional $Z_2\times Z_2'$ symmetry.
Now let consider the ratio $V_{ub}/V_{cb}$ in our mass matrices.
Since $V_{cb}\simeq D_{23}+U^*_{32}$ and 
 $V_{ub}\simeq D_{13}+U^*_{21}D_{23}+U^*_{31}$,
we can express the ratio as
 \begin{equation}         
\left | V_{ub} \o  V_{cb}\right |\simeq \left |{U_{12}^*(U_{23}^* - D_{23})\o
 D_{23}-U_{23}^*} \right |=|U_{12}| \ ,
\end{equation}
\noindent
where we used $U_{31}\simeq U_{12}U_{23}-U_{13}$, $U_{21}\simeq -U_{12}$ and
 $U_{32}\simeq -U_{23}$, and  $D_{13}$ and  $U_{13}$ are neglected.
Thus  the $V_{ub}/V_{cb}$ ratio  depends on only $|U_{12}|$. 
Since we have  $M_{u12}=\l^5$,
which leads to $U_{12}={\cal O}(\l)$, the $V_{ub}/V_{cb}$ ratio is  $a\l$, 
 where $a$ is a  coefficient of ${\cal O}(1)$.
If  $a$ is smaller than $1/2$, our model is consistent with 
the experimental value in  eq.(\ref{CKM}).
The element $M_{u12}=\l^6$,
which leads to $U_{12}={\cal O}(\l^2)$, may be  favored in order to give
  ${V_{ub}/V_{cb}}\simeq \sqrt{m_u/m_c}$.
We cannot get $M_{u12}\simeq \l^6$ in the
 $Z_2\times Z_2'$ symmetry.
 However, since we have $V_{cb}= b\l^2$,  where  a coefficient  $b$ should be fixed 
 to be $0.6-0.7$ with taking $\l=0.22$ by the experimental value
in eq.(\ref{CKM}), we can express $V_{ub}\simeq U_{12}V_{cb}=ab\l^3$.
 Again if $a\leq 1/2$ with  $b=0.6-0.7$,  $V_{ub}\sim \l^4$
is easily attainable. Thus,  $U_{12}={\cal O}(\l)$ is  harmless
as far as $D_{13}$ and $U_{13}$ are suppressed.

In our model, we assumed $\e_1=\e_2=\e_3=\l$.  Even if we take
 another choice such as  $\e_1=\e_2=\l$ and $\e_3=\l^2$,
our conclusion is not so changed.

\section{Summary}
   We have discussed   phenomenological defects of  $U(1)$ 
symmetric quark-lepton mass matrices.
In the quark sector, the mixing $V_{ub}$ is predicted to be  large
compared with the expected value $\l^4$ at the GUT scale.
 In the lepton sector,
     the same $U(1)$ charge such as   $\a_2=\a_3$ is required  to  give 
 the  large mixing in the neutrino sector, however the same charge also leads to
    the large mixing in the charged lepton sector.
   In the viewpoint of the flavor symmetry,
 this is an unpleasant feature because the neutrino mass hierarchy is
  determined only by  unknown   coefficients of ${\cal O}(1)$, and
   the near-maximal  MNS mixing
   is not guaranteed in the case of  both large angle rotations.
  
 To suppress $|V_{ub}|$ below $\l^3$, 
certain entries in the mass matrix  have to be suppressed  
 relative to their naive values.
 This suppression is realized in the additional 
 discrete symmetry $Z_2\times Z_2'$.
This symmetry also leads to the  hierarchical structure
of the neutrino mass matrix  while the large mixing is kept
 in the charged lepton sector.
 Moreover there is  the enhancement of the neutrino mass $m_2$, which
is consistent with the experimental neutrino mass ratio
obtained by atmospheric neutrinos and  solar neutrinos.
 Asking the origin of the $Z_2\times Z_2'$
 symmetry  is a subject for future investigations.

\vskip 1 cm
{\large\bf Acknowledgements}

 The author would like to express  his thanks to the theory  group
of  CFIF/IST in Portugal for their hospitality.
This research is  supported by the Grant-in-Aid for Science Research,
 Ministry of Education, Science and Culture, Japan(No.10140218, No.10640274).  
   

\end{document}